\documentclass[twocolumn,english,superscriptaddress,citeautoscript]{revtex4-2}
\usepackage[T1]{fontenc}
\usepackage[latin9]{inputenc}
\usepackage{array}
\usepackage{multirow}
\usepackage{amsmath}
\usepackage{amssymb}
\usepackage{graphicx}
\usepackage{esint}
\usepackage{xcolor}
\usepackage{tikz}

\usepackage{longtable}           % para tabelas muito grandes
\usepackage{colortbl}            % para cores em tabelas
\usepackage{booktabs}

\makeatletter

\usepackage{wasysym}

\usepackage{babel}

\makeatother

\begin{document}
\title{Nonselective generalized measurements as a resource for quantum thermal machines in a double quantum dot}
\author{Bruno Carvalho}
\affiliation{Department of Physics, Institute of Natural Science, Federal University
of Lavras, 37200-900 Lavras-MG, Brazil}
\author{Jonas F. G. Santos}
\affiliation{Faculty of Exact Sciences and Technology, Federal University
of Grande Dourados, 79804-970 Dourados-MT, Brazil}
\author{Moises Rojas}
\affiliation{Department of Physics, Institute of Natural Science, Federal University
of Lavras, 37200-900 Lavras-MG, Brazil}
\begin{abstract}
We investigated quantum thermal machines powered by sequential nonselective generalized measurements, taking a double quantum dot with coherent interdot tunneling as a working substance. In this platform, the competition between detuning and tunneling hybridizes the localized states and modifies the energetic response of the cycle, allowing us to analyze measurement-driven thermodynamics beyond simple diagonal qubit models. We formulate a three-stroke cycle composed of thermalization with a single reservoir and two generalized measurement channels, and derive the corresponding internal-energy and entropy variations in order to identify the operational regimes of the device. Depending on the measurement parameters, the system can operate as a heat engine, accelerator, heater, or refrigerator. We show that the introduction of tunneling not only reshapes the boundaries between these modes, but also generates refrigeration configurations that are absent in the purely detuned model. In addition, the performance maps reveal that temperature, detuning, and tunneling amplitude jointly control the most favorable regions for work extraction and cooling. Our results demonstrate that coherent interdot coupling acts as an important resource for optimizing measurement-powered quantum thermal machines and highlight double quantum dots as a promising setting for experimentally relevant implementations of measurement-assisted thermodynamic devices.
\end{abstract}
\maketitle

\section{Introduction}

The development of quantum thermodynamics has enabled a fundamental reinterpretation of how work and heat are exchanged at microscopic scales, where both thermal and quantum fluctuations play a central role \cite{vin,kos,camp}. A new class of thermal devices, known as measurement-powered thermal machines, has therefore attracted increasing interest \cite{yi,felce,santos,cao,santos-1,eb,na}. In these systems, the intrinsically invasive nature of quantum measurements is harnessed as an energetic resource, capable of modifying the system's internal energy and driving thermodynamic cycles even in the absence of conventional hot thermal reservoirs. This paradigm highlights the operational role of information and measurement backaction as genuine thermodynamic resources in quantum devices.

Over the past two decades, significant efforts have been devoted to exploring thermodynamic processes in which quantum features of matter play a fundamental role \cite{def,gem}. This line of research has successfully bridged theoretical proposals \cite{Quan,deffner} and experimental realizations \cite{myers,bata}, consolidating the field of quantum thermodynamics and paving the way for practical applications. Among the most prominent systems under investigation are quantum thermal machines, such as the quantum analogues of Carnot, and Otto cycles \cite{kasloff,alicki,mrojas,rojas,mrojas-1}, whose performance has been systematically compared to their classical counterparts in terms of efficiency and irreversibility.
 
Generalized quantum measurements provide an alternative non-equilibrium resource for powering thermodynamics cycles beyond the convencional use of thermal reservoirs and coherent driving. In measurement-powered engines, the measurement back-action associated with positive operator-valued measures (POVMs) can modify both entropy and the internal energy of the working medium, thereby acting as an effective fuel that can later be converted into useful work during subsequent unitary strokes \cite{elouard,klatzow}. Operationally, a distinction must be made between selective and nonselective generalized measurements. In selective measurements, the post-measurement state is conditioned on a particular outcome, requiring either feedback control or postselection; this conditioning can reduce entropy and enables information-to-work conversion protocols. In contrast, nonselective measurements correspond to averaging over all possible outcomes, leading to a completely positive trace-preserving (CPTP) map that typically induces dephasing in the measurement basis while still transferring energy whenever the measurement operators do not commute with the system Hamiltonian \cite{ore,die,be,die-1}. Since generalized measurement operators generally do not commute with the system Hamiltonian, they induce nontrivial energy exchanges during the measurement process. Moreover, any generalized measurement can be decomposed into a sequence of weak measurement \cite{brun}, making the measurement strength a continuously tunable parameter that interpolates between weak and projective regimes. With appropriately engineered measurement channels --either selective or nonselective--such processes can effectively act as thermodynamic resources, functioning as heat-like energy sources that, combined with coherent driving, enable the implementation of consistent measurement-powered quantum thermodynamic cycles \cite{bran,aro,cha}. Furthermore, generalized measurements have been shown to enhance both the charging capacity and ergotropy extraction of quantum batteries, highlighting their role as active control tools in quantum thermodynamics \cite{zh,fra,and}

In this work, we extend the analysis of thermal machines based on generalized measurements to a richer and more realistic physical platform, namely a double quantum dot \cite{haya,mr,mr-1,mr-2}. Unlike the simple qubit models with diagonal Hamiltonians widely explored in the recent literature \cite{vinicius,vinicius-1}, we consider a system in which interdot tunneling introduces coherences and level hybridization, thereby playing a central role in the energetic dynamics and in the conversion of measurement-induced energy into useful thermodynamic tasks. Within this setting, we investigate a three-stroke cycle composed of thermalization with a single thermal reservoir and two sequential nonselective generalized measurements. This framework allows us to identify the corresponding  operational regimes -- including heat engine, accelerator, heater, and refrigerator behavior--and to show that the introduction of tunneling not only reshapes the boundaries between these modes, but also gives rise to refrigeration configurations that are absent in the purely detuned model. In this way, our study highlights coherent interdot coupling as an important resource for optimizing measurement-powered quantum thermal machines.

This paper is structured as follows. In Sec. 2, we describe the model that serves as the working substance.
Section 3 presents the thermodynamic cycle based on nonselective generalized measurements.
In Sec. 4, we analyze the various quantum operational regimes and  their corresponding performance maps. Finally, Sec. 5, contains our conclusions.

\section{Double quantum dot as a working substance }

We consider a double quantum dot (DQD) occupied by a single charge carrier, namely an electron. The electron can be localized either in the left dot, denoted $\left|0\right>$, or in the right dot, denoted by $\left|1\right>$, thus defining an effective two-level system in the localized basis. The energy detuning between the two dots is quantified by the parameter $\varepsilon$, while coherent interdot  tunneling is characterized by the coupling amplitude $\tau$. In this representation, the competition between detuning and tunneling governs the hybridization of the localized states and the resulting energy splitting. 
The Hamiltonian is expressed as
\begin{equation}
H=-\varepsilon\sigma_{z}+\tau\sigma_{x}. \label{eq:hamil}
\end{equation}

The eigenvalues of the Hamiltonian \eqref{eq:hamil}  are
\begin{equation}
\begin{array}{cc}
E_{1}= &\sqrt{\varepsilon^{2}+\tau^{2}}\\
E_{2}= & -\sqrt{\varepsilon^{2}+\tau^{2}},
\end{array}
\end{equation}
where we consider $E_{1}=-E_{2}=E$
and the eigenvectors
\begin{equation}
\begin{array}{cc}
\left|\varphi_{1}\right>= &\cos\theta\left|0\right>+\sin\theta\left|1\right>\\
\left|\varphi_{2}\right>= &\sin\theta\left|0\right>-\cos\theta\left|1\right>,
\end{array}
\end{equation}
with $\theta=\mathrm{arctan}(\frac{\tau}{\sqrt{\varepsilon^{2}+\tau^{2}}-\varepsilon})$.

When weakly coupled to thermal bath at temperature $T$, the system equilibrates to the quantum Gibbs state $\rho(T)=e^{-\beta H}/Z$ with $\beta=1/k_{B}T$. The corresponding partition function is
\begin{equation}
Z(T)=\sum_{i=1}^{2}{ e}^{-\beta E_{i}},\label{eq:Zp}
\end{equation}
where $E_{i}$ are the Hamiltonian eigenvalues, determined by the model parameters.
 
Quantum thermal devices conventionally rely on heat currents between thermal reservoirs. Heat engines extract work from these currents, thermal accelerators amplify the natural energy flow, and refrigerators consume external work to transfer heat from cold bath to a hot one. Here, we propose an alternative architecture in which the thermodynamic cycle is driven by a single thermal reservoir (the cold bath) and two sequentially applied non-selective generalized measurements acting on a two-level realized by a double quantum dot with interdot tunneling.
%%%%%%%%%%

\section{The machine cycle based on generalized measurements}
In this section, we introduce thermodynamic cycle powered by generalized measurement. Such measurements are described by CPTP maps with tunable parameters, which can be implemented through appropriate POVMs. Experimental realizations of related protocols have been reported, for instance, in nuclear magnetic resonance platforms implementing spin quantum heat engines driven by nonselective weak measurements \cite{die-1,vinicius}. In our approach, we consider two sequential nonselective measurement channels with suitably chosen parameters. Their structure is designed to facilitate, for example, the analysis of reversed-order operations and quantum switch protocols \cite{vinicius-1}. The measurement apparatus is modeled as a quantum device with microscopic degrees of freedom coupled to a macroscopic pointer that amplifies the measurement outcome. During the measurement-driven cycle, energy is exchanged between the working substance and distinct meters through their interaction with the system. When the measurement channel modifies the von Neumann entropy of the system, the associated stochastic energy exchange is identified as heat. Conversely, if the channel is isentropic, the exchanged energy is interpreted as work.

\subsection{Nonselective generalized measurement}

The nonselective dynamics associated with each measurement is described by a CPTP map of the form

\begin{equation}
\Phi_{i}(\rho)=\sum_{k=1}^{4}M_{k}^{i}\rho\,M_{k}^{i\dagger},\qquad i\in\{a,b\},
\end{equation}
where $a$ and $b$ label the first and second nonselective measurement channels, respectively. The Kraus operators satisfy the completeness relation 

\begin{equation}
\sum_{k=1}^{4}M_{k}^{i\dagger}M_{k}^{i}=\mathbb{I},
\end{equation}
which ensure trace preservation of the map.
%%%%%%%%%%%%%%%%%%%%%%%%%%%%%%%

The measurement-driven quantum thermal machine operates through a three-stroke cycle.
  
In the first stroke, the working substance thermalizes with a cold reservoir, reaching the Gibbs state $\rho^{(1)}$ at temperature $T$.

 In the second stroke, the system undergoes a generalized nonselective measurement channel controlled by the parameter $a$. For a nonselective measurement, the post-measurement state reads
\begin{equation}
\rho^{(2)}=\sum_{k=1}^{4}M_{k}^{a}\rho^{(1)}\,M_{k}^{a\dagger},
\end{equation} 
 where $M_{k}^{a}$ are the Kraus operators associated  with the first measurement channel.

 In the third stroke, a second nonselective generalized measurement channel characterized by the parameter $b$ is applied, thereby completing the cycle. If $\rho^{(2)}$ denotes the state prior to this measurement, the resulting state becomes
 \begin{equation}
\rho^{(3)}=\sum_{k=1}^{4}M_{k}^{b}\rho^{(2)}\,M_{k}^{b\dagger},
\end{equation} 
where $M_{k}^{b}$ are the Kraus operators of the second measurement channel. Within this framework, the measurement apparatus effectively acts as a thermodynamic resource capable of injecting or extracting energy from the working substance.

Explicitly, the Kraus operators defining the generalized nonselective measurement channels are given as follows.
For the channel characterized by the parameter $a$
\begin{align}
    M_1^a &= \sqrt{1-a}|0\rangle\langle 0|, \nonumber \\ 
    M_2^a &= \sqrt{1-a}|0\rangle\langle 1|, \\
    M_3^a &= \sqrt{a}|1\rangle\langle 1|, \nonumber \\
    M_4^a &= \sqrt{a}|1\rangle\langle 0|,\nonumber 
\end{align}
and for the channel characterized by the parameter $b$,
\begin{align}
    M_1^b &= \sqrt{1-b}|1\rangle\langle 1|, \nonumber \\ 
    M_2^b &= \sqrt{1-b}|1\rangle\langle 0|, \\
    M_3^b &= \sqrt{b}|0\rangle\langle 0|,\nonumber \\
    M_4^b &= \sqrt{b}|0\rangle\langle 1|. \nonumber
\end{align}

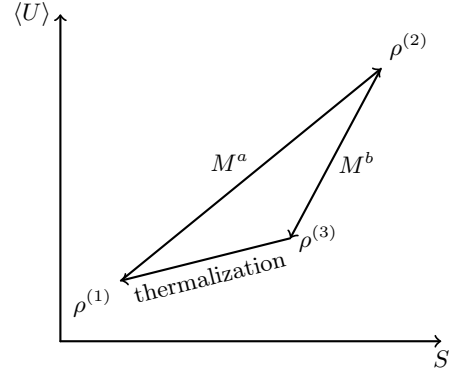
\begin{figure}[t]
\centering
\begin{tikzpicture}[scale=0.8, line cap=round, line join=round]
    \draw[->, thick] (0,0) -- (0,5.4) node[left] {$\langle U\rangle$};
    \draw[->, thick] (0,0) -- (6.3,0) node[below] {$S$};

    \coordinate (r1) at (1.0,1.0);
    \coordinate (r2) at (5.3,4.5);
    \coordinate (r3) at (3.8,1.7);

    \draw[thick,->] (r1) -- (r2);
    \draw[ thick, ->] (r2) -- (r3);
     \draw[thick,->] (r3) -- (r1) node[midway, below, rotate=14] {thermalization};

    \node[below left] at (r1) {$\rho^{(1)}$};
    \node[above right] at (r2) {$\rho^{(2)}$};
    \node[right] at (r3) {$\rho^{(3)}$};
    \node at (2.8,2.95) {$M^{a}$};
    \node at (4.9,2.95) {$M^{b}$};
\end{tikzpicture}
\caption{Schematic cycle in the $S$--$\langle U\rangle$ plane for the measurement-driven machine.}
\label{fig:fig-1}
\end{figure}

%%%%%%%%%%%%%%%%%%%%%%
\subsection{Internal energy and entropy variation in a measurement-driven machine}

To characterize the heat and work in a machine driven by generalized (non-selective) measurements, we derive the energy and entropy changes induced along the cycle. In this framework, the von Neumann entropy and the internal energy of a state $\rho^{(i)}$ are defined as 

\begin{equation}
S^{(i)} = -k_B \mathrm{Tr}\left [\rho^{(i)} \ln \rho^{(i)}\right ],
\end{equation}
\begin{equation}
U^{(i)} =  \mathrm{Tr}\left [\mathcal{H} \rho^{(i)} \right ].
\end{equation}
Here, $i=1,2,3$ labels the successive stages of the cycle. The internal-energy variations at each stroke read

\begin{align}
    \left\langle \Delta U^{(i)} \right\rangle &=  \left\langle  U^{(i)}- U^{(i-1)}\right\rangle \nonumber \\ 
     &= \mathrm{Tr}\left [\mathcal{H}(\rho^{(i)}-\rho^{(i-1)})\right ], \nonumber \\
\end{align}
while the entropy changes are
\begin{align}
    \Delta S_1 &= S^{(1)}-S^{(3)}, \nonumber \\ 
    \Delta S_2 &= S^{(2)}-S^{(1)}, \\
    \Delta S_3 &= S^{(3)}-S^{(2)}. \nonumber 
\end{align}

The initial state is the thermal equilibrium state

\begin{equation}
\rho^{(1)} =\frac{e^{-\beta E}}{Z}\left|\varphi_{1}\right>\langle \varphi_{1}|+\frac{e^{\beta E}}{Z}\left|\varphi_{2}\right>\langle \varphi_{2}|,
\end{equation}
with partition function $Z=e^{-\beta E}+e^{\beta E}$. After the first measurements, the state becomes

\begin{equation}
\rho^{(2)} = (1-a) \left|0\right>\langle 0|+a\left|1\right>\langle 1|,
\end{equation}
and, after second measurement channel, the density operator becomes

\begin{equation}
\rho^{(3)} = b \left|0\right>\langle 0|+(1-b)\left|1\right>\langle 1|.
\end{equation}

For this model, the internal energy variations along the three strokes are
 
\begin{align} 
   \left\langle \Delta U_1 \right\rangle &= -E \tanh(\beta E)+\varepsilon (2b-1), \nonumber \\
    \left\langle  \Delta U_2 \right\rangle&= E \tanh(\beta E)+\varepsilon (2a-1),  \\
    \left\langle  \Delta U_3 \right\rangle&= 2 \varepsilon (1-a-b), \nonumber \label{eq:delta}
\end{align}

whereas the corresponding entropy changes read

 \begin{align} 
    \Delta S_1 &= k_{B} \left [h\left (\frac{1}{2}(1-\tanh(\beta E))\right )- h(b)\right ], \nonumber \\
     \Delta S_2 &= k_{B} \left [h\left (a\right )-h\left (\frac{1}{2}(1-\tanh(\beta E))\right )\right ],  \\
      \Delta S_3 &= k_{B} \left [h\left (b\right )-h\left (a\right )\right ], \nonumber
\end{align}
with $h(u)=-u \ln(u)-(1-u)\ln(1-u)$ the binary Shannon entropy.
 
These results provide a clear thermodynamic interpretation of each stroke. Processes with finite entropy production $(\Delta S_{i}\neq 0)$ are associate with irreversible, measurements-induced heat exchange, whereas nearly isentropic strokes correspond to work-like energy transfer arising from measurement backaction.

 %%%%%%%%%%%%%%
\begin{table}[htbp]
    \centering
    \begin{tabular}{ccccc}
   % \begin{tabular}{|c|c|c|c|c|}
   \toprule
    %\hline
    \textbf{Operation Mode} & $Q_{h}$ & $Q_{c}$ & $W$ & Thermal efficiency \\
    \midrule
   % \hline
    %\hline
    Engine & $+$ & $-$ & $-$ & $\eta = \left|\frac{W}{Q_h}\right|$\\ 
    Refrigerator & $-$ & $+$ & $+$ & $\text{CoP}= \left|\frac{Q_c}{W}\right|$ \\ 
    Accelerator & $+$ & $-$ & $+$ & $\text{CoP} = \left|\frac{Q_h}{W}\right|$\\ 
    Heater & $-$ & $-$ & $+$ & $\text{CoP} = \left|\frac{Q_h}{W}\right|$\\
    \bottomrule
    %\hline
    \end{tabular}
    \caption{Classification of the operational modes of a thermal machine}
    \label{tab:clas}
\end{table}

\subsection{Thermal efficiency}

In Table~\ref{tab:clas}, we present a classification of the different operational modes of a thermal machine. In the engine regime, the efficiency  $\eta$ satisfies  $0 \leq \eta < 1$. For the refrigerator, heater, and accelerator modes, the relevant quantity is the coefficient of performance
$(COP)$.
Since the $COP$ is unbounded and may diverge near regime transitions, we define the normalized parameter
\begin{equation}
\kappa=\frac{COP}{1+COP}.\label{eq:kp}
\end{equation}
which maps $0<COP < \infty$ onto $0<\ensuremath{\kappa}<1$. Here, $\kappa \to 0$
indicates poor performance, while $\kappa \to 1$ corresponds to optimal operation (with $COP=1$ yielding $\kappa=0.5$). This normalization provides a unified bounded scale to compare all operational modes.

\begin{figure}[t]
\centering
\begin{tikzpicture}[scale=0.8, line cap=round, line join=round, >=stealth]
    % top-left
    \begin{scope}[shift={(0,0)}]
    \node[above] at (2.15,4.4) {\textbf{engine}};
        \draw[->, thick] (0,0) -- (0,4.4) node[above] {$\langle U\rangle$};
        \draw[->, thick] (0,0) -- (4.3,0) node[right] {$S$};
        \coordinate (r1) at (0.9,0.9);
        \coordinate (r2) at (3.6,3.4);
        \coordinate (r3) at (3.6,1.3);
        \draw[thick,->] (r1) -- (r2) node[midway, above left] {$M^{a}$};
        \draw[thick,->] (r2) -- (r3) node[midway, right] {$M^{b}$};
        \draw[thick,->] (r3) -- (r1) node[midway, below, rotate=8
       ] {thermalization};
        \node[left] at (r1) {$\rho^{(1)}$};
        \node[right] at (r2) {$\rho^{(2)}$};
        \node[right] at (r3) {$\rho^{(3)}$};
    \end{scope}

 \begin{scope}[shift={(5.6,0)}]
 \node[above] at (2.15,4.4) {\textbf{accelerator}};
        \draw[->, thick] (0,0) -- (0,4.4) node[above] {$\langle U\rangle$};
        \draw[->, thick] (0,0) -- (4.3,0) node[right] {$S$};
        \coordinate (r1) at (0.9,0.9);
        \coordinate (r2) at (3.5,1.3);
        \coordinate (r3) at (3.5,3.4);
        \draw[thick,->] (r1) -- (r2) node[midway, below] {$M^{a}$};
        \draw[thick,->] (r2) -- (r3) node[midway, right] {$M^{b}$};
         \draw[thick,->] (r3) -- (r1) node[midway, above, rotate=44] {thermalization};
        \node[left] at (r1) {$\rho^{(1)}$};
        \node[right] at (r2) {$\rho^{(2)}$};
        \node[right] at (r3) {$\rho^{(3)}$};
    \end{scope}

    % bottom-center
    \begin{scope}[shift={(3.1,-5.5)}]
    \node[above] at (2.25,4.2) {\textbf{heater}};
        \draw[->, thick] (0,0) -- (0,4.2) node[above] {$\langle U\rangle$};
        \draw[->, thick] (0,0) -- (4.5,0) node[right] {$S$};
        \coordinate (r1) at (1.2,2.1);
        \coordinate (r2) at (3.6,0.9);
        \coordinate (r3) at (3.6,3.5);
        \draw[thick,->] (r1) -- (r2) node[midway, below left] {$M^{a}$};
        \draw[thick,->] (r2) -- (r3) node[midway, right] {$M^{b}$};
        \draw[thick,->] (r3) -- (r1) node[midway, above, rotate=31] {thermalization};
        \node[left] at (r1) {$\rho^{(1)}$};
        \node[right] at (r2) {$\rho^{(2)}$};
        \node[right] at (r3) {$\rho^{(3)}$};
    \end{scope}
\end{tikzpicture}
\caption{Schematic cycles in the $S$--$\langle U\rangle$ plane for the measurement-driven machine, showing the three possible orderings of the measurement-induced strokes.}
\label{fig:fig-2}
\end{figure}
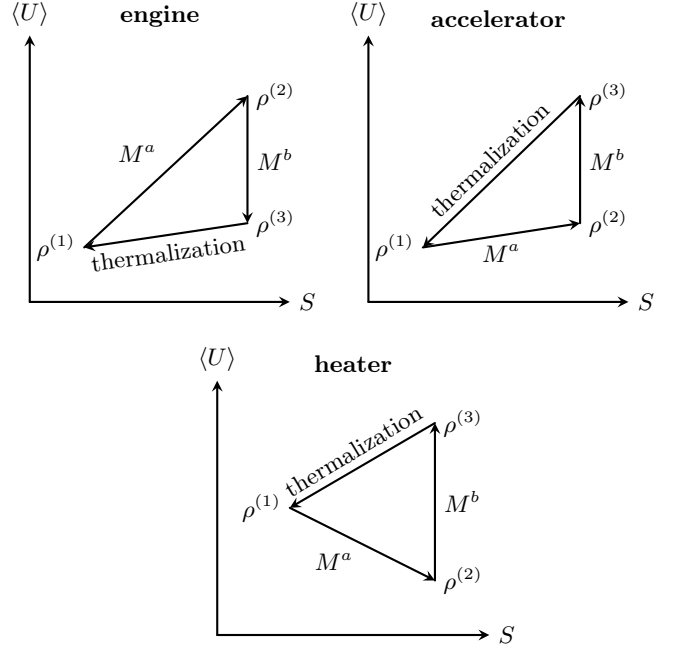

\begin{figure*}[t] % [t] posiciona a figura no topo da página
    \centering
    \includegraphics[width=1.2\columnwidth]{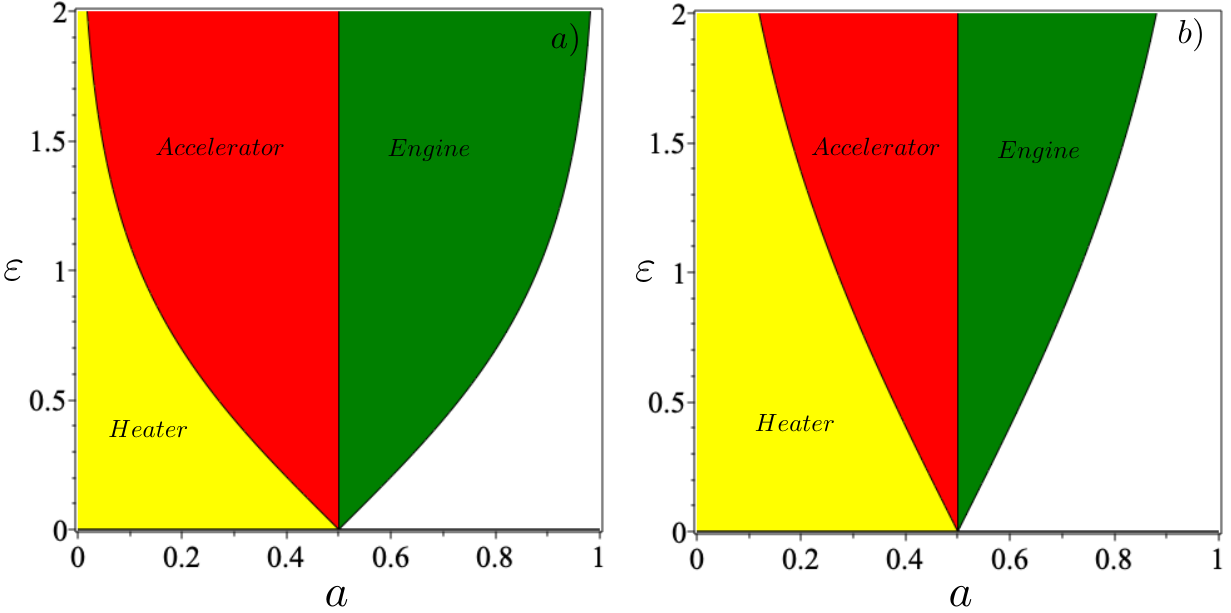} % Substitua pelo nome do arquivo da imagem
    \caption{The operational modes of the quantum cycle as a function of $a$ and $\varepsilon$ for different parameter configurations. a) For  $T=1$. b) For $T=2$. The regions in the plots correspond to distinct operational regimes: engine (green),
heater (yellow), and accelerator (red). The parameter $\tau$ was fixed at $\tau=0$.}
    \label{fig:gene-1}
\end{figure*}

\begin{figure*}[t] % [t] posiciona a figura no topo da página
    \centering
    \includegraphics[width=1.2\columnwidth]{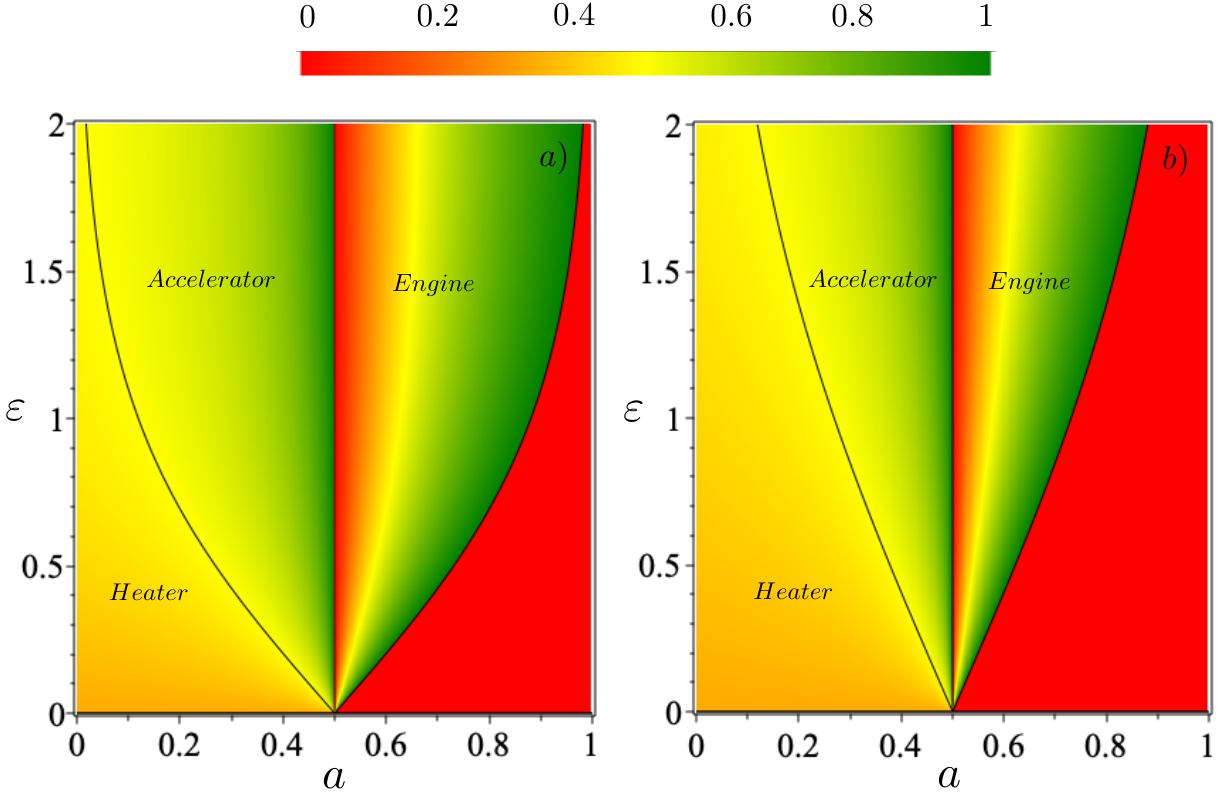} % Substitua pelo nome do arquivo da imagem
    \caption{Performance of the quantum machine based on generalized measurement as a function of $a$ and $\varepsilon$ for different parameter configurations. a) For $T=1$. b) For $T=2$. The parameter $\tau$ was fixed at $\tau=0$.}
    \label{fig:gene-eff-1}
\end{figure*}

%%%%%%%%

\section{Quantum operational regimes}

In this section, we analyze the operational regimes of the measurement-driven quantum thermal device introduced above. Although the cycle is powered by two nonselective generalized measurements and a single thermal reservoir, the interplay between measurement backaction and thermalization gives rise to qualitatively distinct thermodynamic behaviors. By tuning the measurement parameters $a$ and $b$, the device can operate in different modes, including a heat engine, a refrigerator, and other dissipative regimes such as heaters and thermal accelerators. These regimes are identified through the signs of the total work performed over the cycle and the net heat exchanged with the thermal bath. Figure \ref{fig:fig-1} summarizes the phase diagram of the machine, highlighting the regions in parameter space associated with each operational mode. The emergence of these regimes demonstrates that generalized measurements can effectively replace conventional thermal reservoirs, acting as controllable thermodynamic resources capable of driving energy conversion processes at the quantum level.

\begin{figure*}[t] % [t] posiciona a figura no topo da página
    \centering
    \includegraphics[width=1.2\columnwidth]{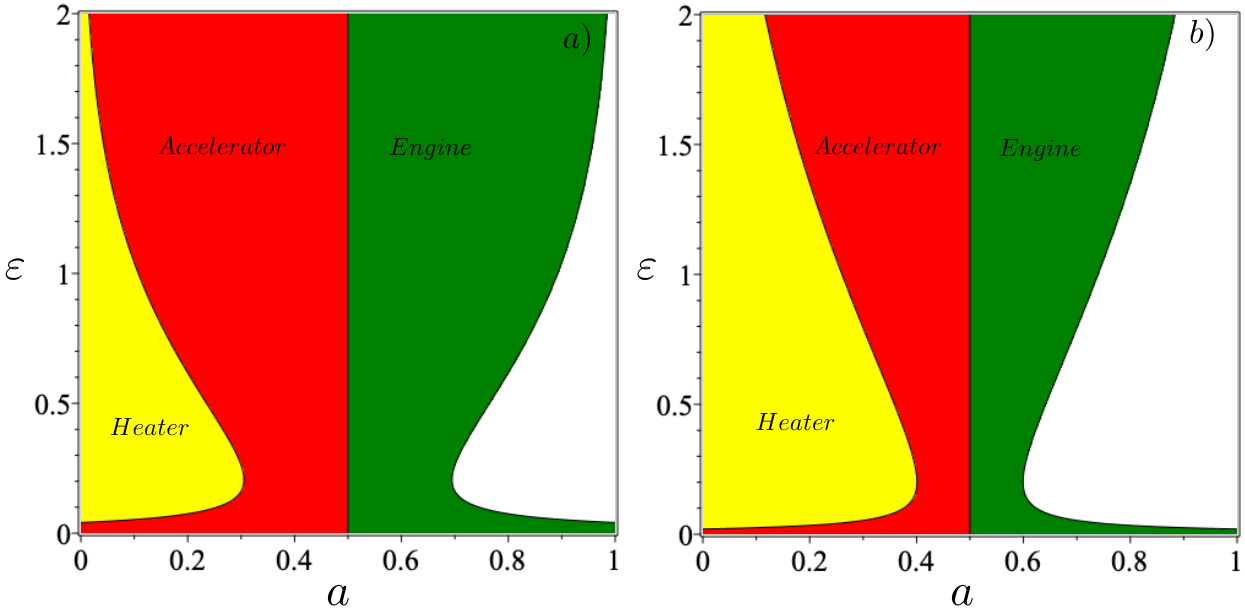} % Substitua pelo nome do arquivo da imagem
    \caption{The operational modes of the quantum cycle as a function of $a$ and $\varepsilon$ for different parameter configurations. a) For  $T=1$. b) For $T=2$. The regions in the plots correspond to distinct operational regimes: engine (green),
heater (yellow), and accelerator (red). The parameter $\tau$ was fixed at $\tau=0.2$.}
    \label{fig:engine-1}
\end{figure*}

\begin{figure*}[t] % [t] posiciona a figura no topo da página
    \centering
    \includegraphics[width=1.2\columnwidth]{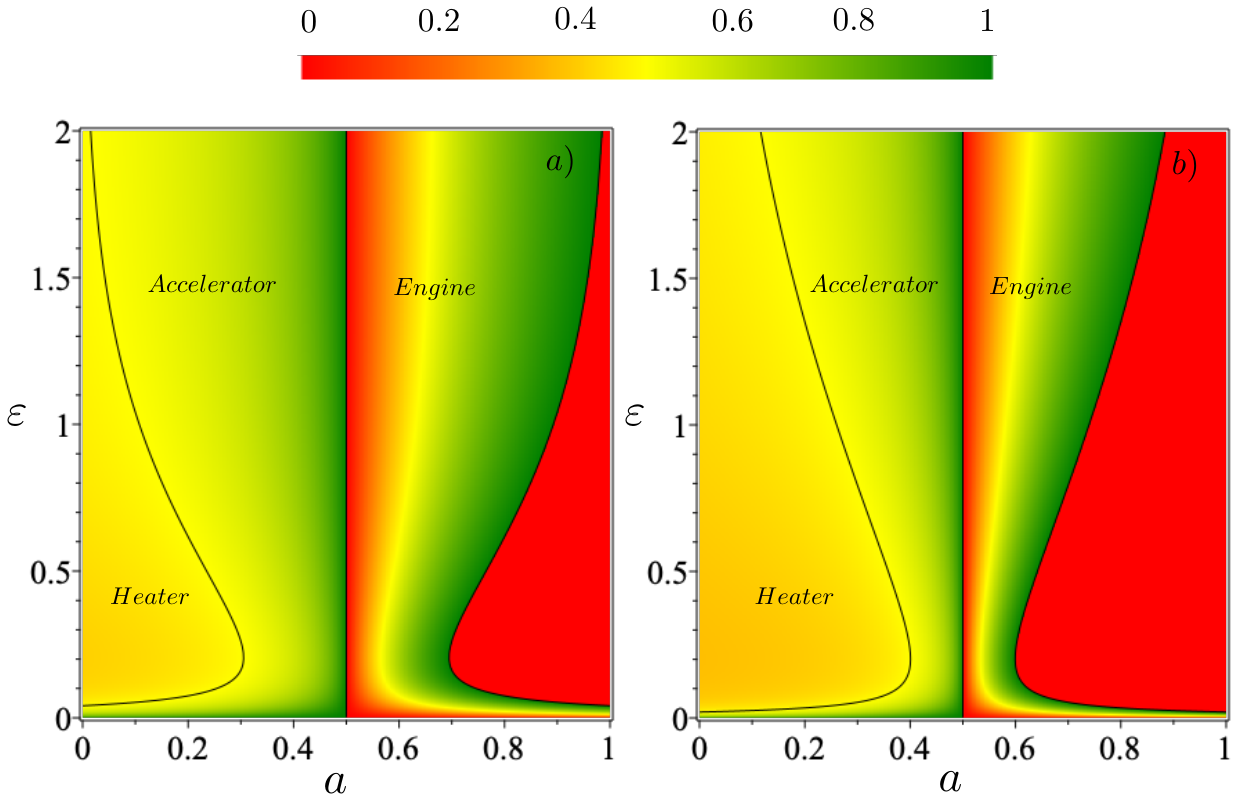} % Substitua pelo nome do arquivo da imagem
    \caption{Performance of the measurement-driven quantum machine in the  $(a,\varepsilon)$ plane for $\tau=0.2$ at a) $T=1$ and b) $T=2$.}
    \label{fig:engine-eff-1}
\end{figure*}

\begin{figure}[t]
\centering
%\resizebox{\columnwidth}{!}{%
\begin{tikzpicture}[line cap=round, line join=round, >=stealth, font=\small]
    \begin{scope}[shift={(0,0)}]
     \node[above] at (2.0,4.4) {\textbf{refrigerator}};
        \draw[->, thick] (0,0) -- (0,4.4) node[above] {$\langle U\rangle$};
        \draw[->, thick] (0,0) -- (3.8,0) node[right] {$S$};
        \coordinate (r1) at (0.9,1.3);
        \coordinate (r2) at (0.9,3.4);
        \coordinate (r3) at (3.5,0.8);
        \draw[thick,->] (r1) -- (r2) node[midway, left] {$M^{a}$};
        \draw[thick,->] (r2) -- (r3) node[midway, above right] {$M^{b}$};
        \draw[thick,->] (r3) -- (r1) node[midway, below, rotate=350] {thermalization};
        \node[left] at (r1) {$\rho^{(1)}$};
        \node[left] at (r2) {$\rho^{(2)}$};
        \node[right] at (r3) {$\rho^{(3)}$};
    \end{scope}
    \begin{scope}[shift={(4.6,0)}]
     \node[above] at (2.3,4.4) {\textbf{accelerator}};
        \draw[->, thick] (0,0) -- (0,4.1) node[above] {$\langle U\rangle$};
        \draw[->, thick] (0,0) -- (3.6,0) node[right] {$S$};
\coordinate (r1) at (0.9,0.9);
        \coordinate (r2) at (0.9,2.4);
        \coordinate (r3) at (3.1,3.3);
        \draw[thick,->] (r1) -- (r2) node[midway, left] {$M^{a}$};
        \draw[thick,->] (r2) -- (r3) node[midway, above left] {$M^{b}$};
        \draw[thick,->] (r3) -- (r1) node[midway, below, rotate=47] {thermalization};
        \node[left] at (r1) {$\rho^{(1)}$};
        \node[left] at (r2) {$\rho^{(2)}$};
        \node[right] at (r3) {$\rho^{(3)}$};
    \end{scope}
    \begin{scope}[shift={(2.3,-5.5)}]
    \node[above]at (2.25,4.2){\textbf{heater}};
        \draw[->, thick] (0,0) -- (0,4.2) node[above] {$\langle U\rangle$};
        \draw[->, thick] (0,0) -- (4.5,0) node[right] {$S$};
        \coordinate (r1) at (1.2,1.0);
        \coordinate (r2) at (1.2,3.4);
        \coordinate (r3) at (3.2,2.1);
        \draw[thick,->] (r1) -- (r2) node[midway, left] {$M^{a}$};
        \draw[thick,->] (r2) -- (r3) node[midway, above right] {$M^{b}$};
       \draw[thick,->] (r3) -- (r1) node[midway, below, rotate=29] {thermalization};
        \node[left] at (r1) {$\rho^{(1)}$};
        \node[left] at (r2) {$\rho^{(2)}$};
        \node[right] at (r3) {$\rho^{(3)}$};
    \end{scope}
   % \label{fig:engine-diag}
\end{tikzpicture}%
%\label{fig:engine-diag}}
\caption{Schematic refrigeration cycles in the $S$--$\langle U\rangle$ plane. Depending on the ordering of $M^{a}$, $M^{b}$, and thermalization, the cycle realizes different refrigeration-related trajectories.}
\label{fig:refri-scheme}
\end{figure}
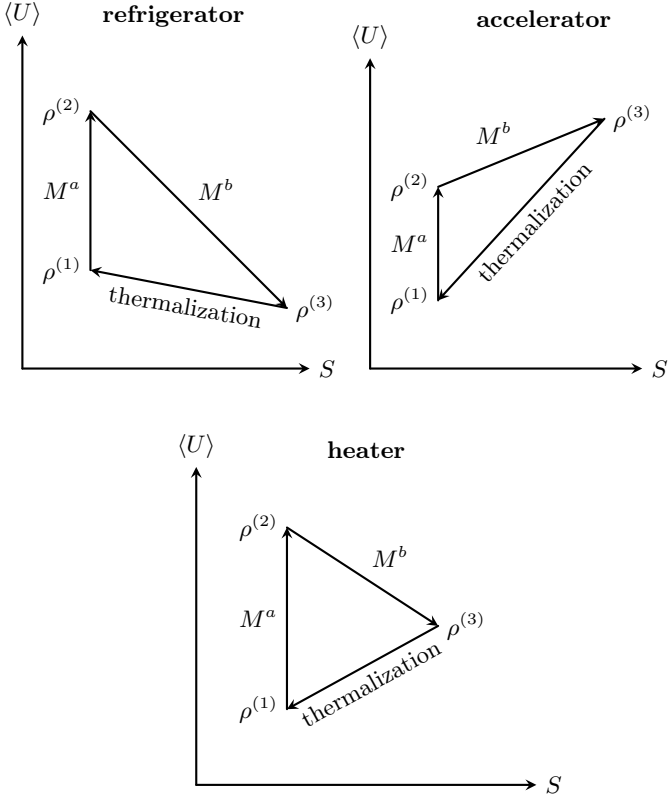

\begin{figure*}[t] % [t] posiciona a figura no topo da página
    \centering
    \includegraphics[width=1.2\columnwidth]{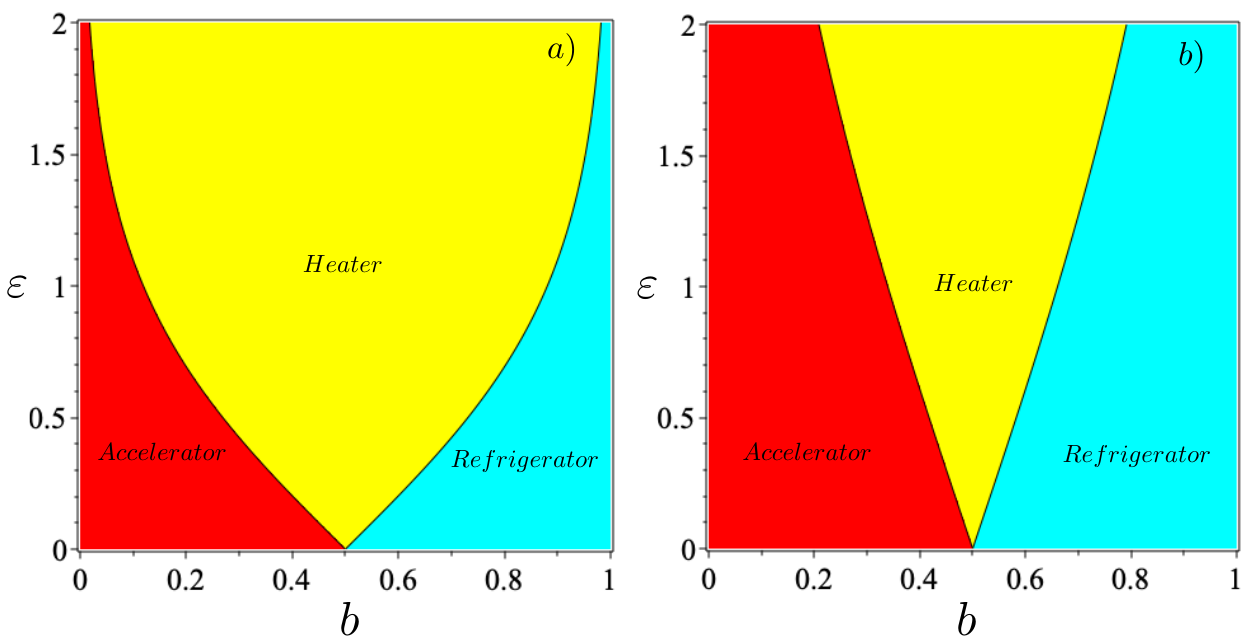} % Substitua pelo nome do arquivo da imagem
    \caption{Operational modes of the quantum cycle driven by generalized measurements as a function of $b$ and $\varepsilon$ for different parameter configurations. a) For $T=1$. b) For $T=3$. The parameter $\tau$ was fixed at $\tau=0$. The regions in the plots correspond to distinct operational regimes: refrigerator (cyan),
heater (yellow), and accelerator (red).}
    \label{fig:refri-1(zero)}
\end{figure*}
\begin{figure*}[t] % [t] posiciona a figura no topo da página
    \centering
    \includegraphics[width=1.2\columnwidth]{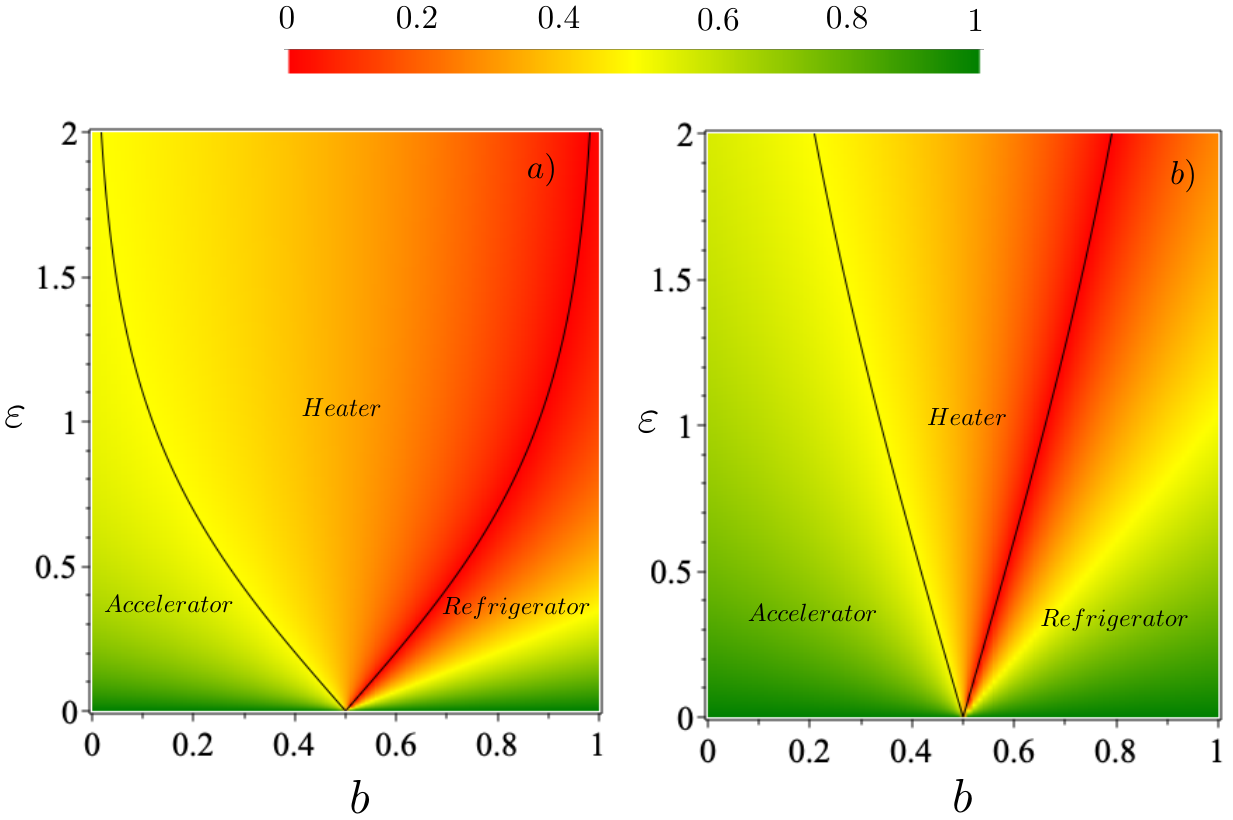} % Substitua pelo nome do arquivo da imagem
    \caption{Performance of the measurement-driven quantum machine in the  $(b,\varepsilon)$ plane for $\tau=0$ at a) $T=1$ and b) $T=3$.}
    \label{fig:refri-eff-1(zero)}
\end{figure*}

%%%%%%%%%%%%%%%%%%%%%%%%

\subsection{Quantum engine}

The quantum heat engine regime is characterized by the net extraction of work from the system, powered by the energy injected during the nonselective measurement stroke. In this configuration, the first measurement channel $M_{a}$ provides the effective quantum heat $Q_{h}$, promoting population redistribution and increasing both the internal energy and entropy of working substance. In the subsequent stroke, the second measurement channel $M_{b}$ enables the extraction of work, effectively converting part of the injected energy into useful output. The final thermalization step restores the system to equilibrium, closing the cycle. For the device to operate as a heat engine, the third stroke must be isentropic, i.e. $\Delta S_{3}=0$. This condition implies $h(b)=h(a)$, leading to two possible solutions: $b=a$ or $b=1-a$. However, the latter corresponds to a trivial situation in which no net work is extracted. Therefore, we focus on the physically relevant case $b=a$.
Under this condition, the thermodynamic quantities simplify to
 \begin{align}
 &Q_c=\left\langle \Delta U_1\right\rangle=-E\tanh{\left(\beta E\right)}+\varepsilon(2a-1),\\
&Q_h=\left\langle\Delta U_2\right\rangle=E\tanh{\left(\beta E\right)}+\varepsilon(2a-1),\\
&W=\left\langle\Delta U_3\right\rangle=2\varepsilon(1-2a).\\ \nonumber
%&\Delta U_2 > 0 \iff a>\frac{1}{2}\left[1-\frac{E}{\varepsilon}\tanh{\left(\beta E\right)}\right]\\
%&\Delta U_1 < 0 \iff a<\frac{1}{2}\left[1+\frac{E}{\varepsilon}\tanh{\left(\beta E\right)}\right]\nonumber
\end{align}
The engine operates when the heat absorbed from the measurement channel is positive and the heat released  to the bath is negative, namely $Q_{h}>0$ and $Q_{c}<0$, while the work $W<0$. These conditions impose constrains on the measurement strength $a$, given by $\frac{1}{2}\leqslant a<\frac{1}{2}\left[1+\frac{E}{\varepsilon}\tanh{\left(\beta E\right)}\right]$.
 %\begin{align}
%& \frac{1}{2}<a<\frac{1}{2}\left[1+\frac{E}{\varepsilon}\tanh{\left(\beta E\right)}\right].
%\end{align}
It is worth noting that the heat exchanged during the thermalization process, $\Delta U_{1}$, is associated with the cold reservoir, while the heat $\Delta U_{2}$ originates from the first  nonselective generalized measurement. This clearly highlights the role of measurement backaction as an effective hot source driving the engine cycle. Beyond the engine regime, the device may also operate in other dissipative modes, such as thermal accelerator and heater, as shown in Fig. \ref{fig:fig-2}. 

In the accelerator regime, the second stroke injects heat $Q_{h}>0$ into the working substance via the measurement $M^{a}$, while work $W>0$ is performed on the system during the third stroke. The cycle is completed by thermalization, during which heat is released to the bath $Q_{c}<0$. In this case, the parameter $a$ satisfies $\frac{1}{2}\left[1-\frac{E}{\varepsilon}\tanh{\left(\beta E\right)}\right]<a<\frac{1}{2}$. 

On the other hand, for the heater regime, the system first thermalizes with the cold bath. Subsequently, the measurement apparatus absorbs energy during the second stroke, effectively acting as a heater sink, while the third measurement channel supplies work to the system. Here, the parameter $a$ obeys $a<\frac{1}{2}\left[1-\frac{E}{\varepsilon}\tanh{\left(\beta E\right)}\right]$. These regimes further illustrate the versatility of measurement-driven quantum devices, where the same physical setup can realize qualitatively distinct thermodynamic behaviors depending on the measurement parameters.

%\subsection{Quantum engine}  
  
Figure \ref{fig:gene-1} shows the operational regimes of the generalized-measurement quantum cycle in the $(a,\varepsilon)$ plane for $\tau=0$ \cite{vinicius}. In panel (a), corresponding to $T=1$, three distinct regions are observed. For small values of $a$ and low detuning $\varepsilon$, the device operates in the heater mode, whereas increasing $\varepsilon$ at the same small $a$ drives the system into the accelerator regime. For $a\geqslant 0.5$, the cycle operates as a heat engine, and this engine region broadens as the detuning increases. The white region denotes parameter sets where the heat and work fail to satisfy the sign conventions of any well-defined operational mode, and the cycle therefore lacks a consistent thermodynamic behavior. In panel (b), for $T=2$, the overall structure is preserved, but the boundary between heater and accelerator is displaced upward, so that the heater region expands toward larger values of $\varepsilon$ while the accelerator region is correspondingly reduced. By contrast, the engine domain remains concentrated in the $a \geqslant 0.5$ sector, showing that increasing temperature mainly affects the competition between heater and accelerator operation for $a<0.5$. The white region corresponds to parameter for which the exchanged heat and work do not satisfy the sign conditions required for any well-defined operational mode. In that sector, the cycle does not realize a consistent engine, accelerator, or heater behavior. 
 
In Fig.  \ref{fig:gene-eff-1}, we present the performance map of the same cycle in the $(a,\varepsilon)$ parameter space. In panel (a), for $T=1$, the best performance is found inside the engine region away from the transition lines, while the vicinity of the boundaries between operational modes is associated with lower performance. The accelerator sector displays intermediate values of the normalized performance, whereas the heater region is characterized by comparatively poorer performance, especially at small detuning. In panel (b), for $T=2$, the high-performance sector onside the engine regime remains robust, but the redistribution of the boundaries modifies the low-performance zones near the heater--accelerator transition. Overall, Fig. \ref{fig:gene-eff-1} indicates that the engine mode provides the most favorable thermodynamic response in this branch, whereas the heater and accelerator modes exhibit more modest performance and stronger sensitivity to the detuning and measurement strength.

In Fig. \ref{fig:engine-1}, we shows how the operational modes of the engine branch are modified when a finite tunneling amplitude $\tau=0.2$ is introduced. In the panel (a), for $T=1$, the phase diagram still exhibits the same three regimes observed in the zero-tunneling case, namely heater, accelerator, and engine, but their boundaries are noticeably distorted by the coherent hybridization between the localized states. The heater region remains confined to small values of $a$ and low detuning $\varepsilon$, while the accelerator regime occupies most of the sector $a<0.5$ for intermediate and large detunings. For $a>0.5$, the engine mode remains dominant, indicating that the work-producing regime is robust agains moderate tunneling. However, the right-hand boundary of the engine region is no longer a simple monotonic curve: at very large values of $a$ and small detuning, a narrow non-engine sector appears, showing the excessively strong measurements may suppress useful work extraction when the level splitting is weak.  
In panel (b), corresponding to $T=2$, the same qualitative structure is preserved, but the competition between the three regimes is qualitatively reshaped by the higher temperature. The heater region expands toward larger values of $\varepsilon$ for small $a$, indicating that thermal fluctuations facilitate dissipative behavior in this part of parameter space. As a consequence, the accelerator region is slightly compressed, although it still occupies most of the interval below $a=0.5$. The engine regime continues to dominate the sector $a>0.5$, confirming that finite tunneling does not destroy the work-producing branch. Nevertheless, the white low-performance region at large $a$ and small detuning becomes more visible, indicating that increasing temperature enlarges the parameter window in which the measurement-driven cycle ceases to operate efficiently as an engine. In summary, Fig. \ref{fig:engine-1} shows that finite tunneling preserves the main structure of the engine branch, introducing only quantitative changes in the boundaries between heater, accelerator, and engine operational mode. 

Figure \ref{fig:engine-eff-1} displays the thermal-performance maps for the engine branch in the presence of finite tunneling amplitude $\tau=0.2$. In panel (a), for $T=1$, the highest values of the normalized efficiency are concentrated within the engine region, particularly at intermediate and large values of $a$, away from the boundaries separating it from the accelerator regime. This indicates that, once the measurement strength is sufficient  to place the cycle in the work-producing regime,  the conversion of measurement-induced energy into useful work becomes more effective. In contrast, the low-performance region observed at large $a$ and small detuning $\varepsilon$ shows that very strong measurements do not necessarily enhance engine operation; when the level splitting is small, measurement backaction tends to increase irreversibility. In panel (b), for $T=2$, the overall structure is preserved, but the performance landscape is quantitatively modified. The high-performance region within the engine sector becomes more localized and shifts slightly with increasing $\varepsilon$, indicating that thermal fluctuations affect the efficiency of work extraction. At the same time, increasing temperature reduces the performance of the accelerator regime, while the low-performance heater region expands.

\subsection{Quantum refrigerator}

The quantum refrigerator regime is characterized by the extraction of heat from the thermal reservoir, driven by the work supplied through the nonselective measurement processes. In this configuration, the measurement channels act as external resources that inject energy into the system, enabling a cooling cycle even in the presence of a single thermal bath. In the first stroke, the measurement channel $M^{a}$ modifies the state of the working substance, redistributing the population and generally increasing its internal energy. This process effectively plays the role of a work input. In the subsequent stroke, the second measurement channel $M^{b}$ further drives the system away from the equilibrium, preparing it in a nonequilibrium state prior to thermalization. During the final stroke, the system relaxes back to the Gibbs state through contact with the thermal bath. As a result of the previous measurement-induced transformations, heat is extracted from the reservoir during this relaxation process, thereby realizing a refrigeration cycle. For the device to operate as a refrigerator, the second stroke must be isentropic, i.e, $\Delta S_{2}=0$. This condition imposes $h(a)=h(\frac{1}{2}\left[1- \tanh{\left(\beta E\right)} \right])$, yielding the two possible solutions
\begin{equation}
a=\frac{1}{2}\left[1\pm\tanh\left(\beta E\right)\right].
\end{equation}
Under this condition, the energy exchanged during this stage is purely work-like, ensuring that the entropy reduction required for cooling is achieved during the thermalization stage. The operational modes of the cycle refrigerador, accelerator, and heater are presented in Fig. \ref{fig:refri-scheme}.
%%%%%%%
\subsubsection{Branch $a = \frac{1}{2}\left[1 + \tanh(\beta E)\right]$}
Focusing on the branch
\begin{equation}
a=\frac{1}{2}\left[1+\tanh\left(\beta E\right)\right],
\end{equation}
and using Eq. (\ref{eq:delta}), the thermodynamic quantities take the form
 \begin{align}
 &Q_c=\left\langle\Delta U_1\right\rangle=-E\tanh{\left(\beta E\right)}+\varepsilon(2b-1). \nonumber \\
&W=\left\langle\Delta U_2\right\rangle=(E+\varepsilon)\tanh{\left(\beta E\right)}.\\ 
&Q_h=\left\langle\Delta U_3\right\rangle=\varepsilon(1-\tanh{\left(\beta E\right)}-2b). \nonumber
\end{align} 
The refrigerator regime requires heat extraction from the thermal bath, and heat delivered to the measurement channel $M^{b}$ namely $Q_{c}>0$, $W>0$ and $Q_{h}<0$. Since $W>0$, the operating condition is determined by the signs of $Q_{c}$ and $Q_{h}$, leading to $b>\frac{1}{2}\left[1+\frac{E}{\varepsilon}\tanh\left(\beta E\right)\right ]$.

In addition to the refrigerator regime, the device can operate in the other thermodynamic modes. The accelerator regime is characterized by $Q_{c}<0$, $W>0$ and $Q_{h}>0$, which yields the condition $b<\frac{1}{2}\left[1-\tanh\left(\beta E\right)\right ]$.

The heater regime corresponds to $Q_{c}<0$, $W>0$ and $Q_{h}<0$, defining the intermediate region $\frac{1}{2}\left[1-\tanh\left(\beta E\right)\right ]<b<\frac{1}{2}\left[1+\frac{E}{\varepsilon}\tanh\left(\beta E\right)\right ]$.

\begin{figure*}[t] % [t] posiciona a figura no topo da página
    \centering
    \includegraphics[width=1.2\columnwidth]{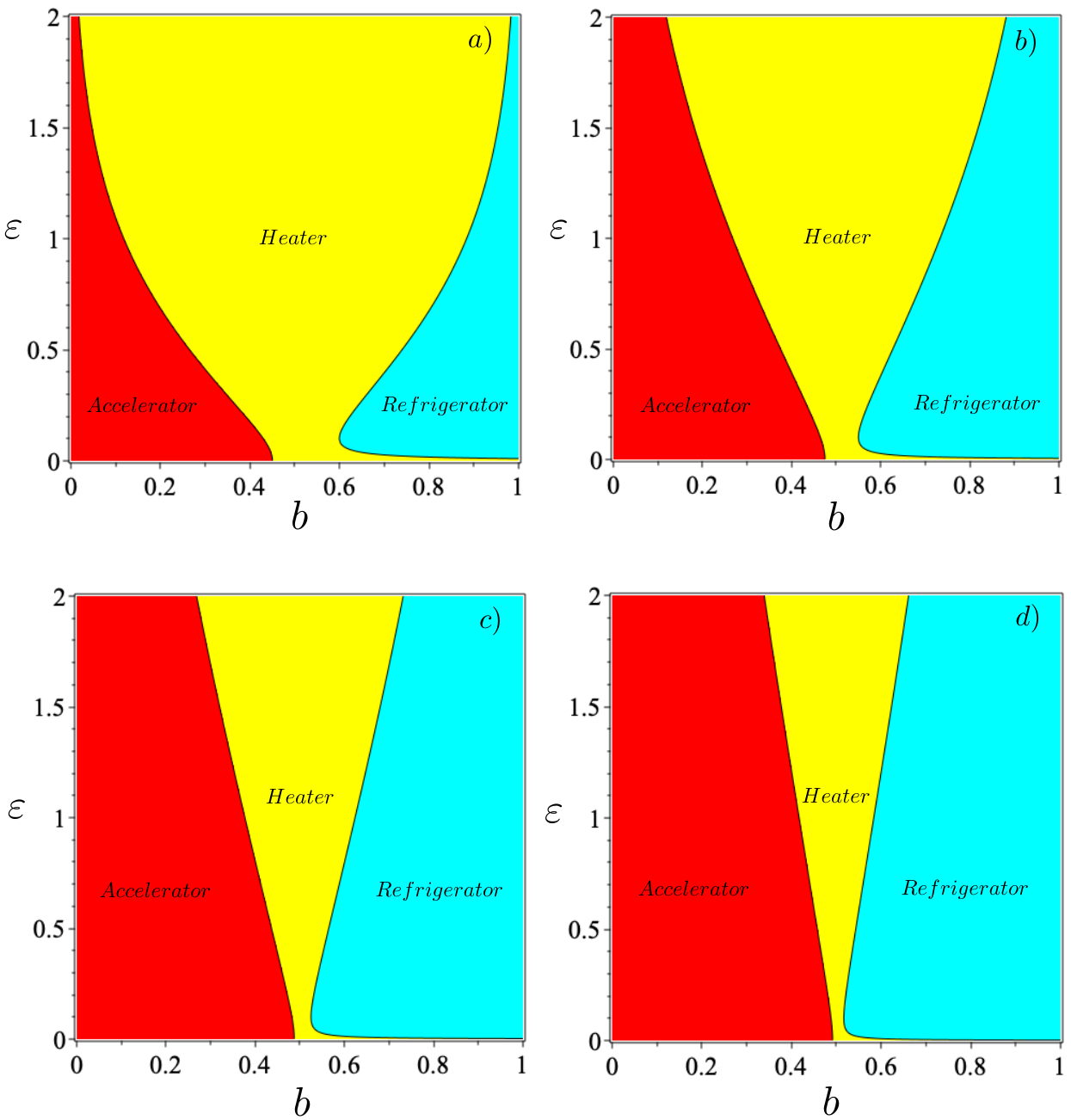} % Substitua pelo nome do arquivo da imagem
    \caption{The operational modes of the quantum cycle as a function of $b$ and $\varepsilon$ for different parameter configurations. a) For  $T=1$. b) For $T=2$. c) For $T=4$. d) For $T=6$. The parameter $\tau$ was fixed at $\tau=0.1$.}
    \label{fig:refri-1}
\end{figure*}

\begin{figure*}[t] % [t] posiciona a figura no topo da página
    \centering
    \includegraphics[width=1.2\columnwidth]{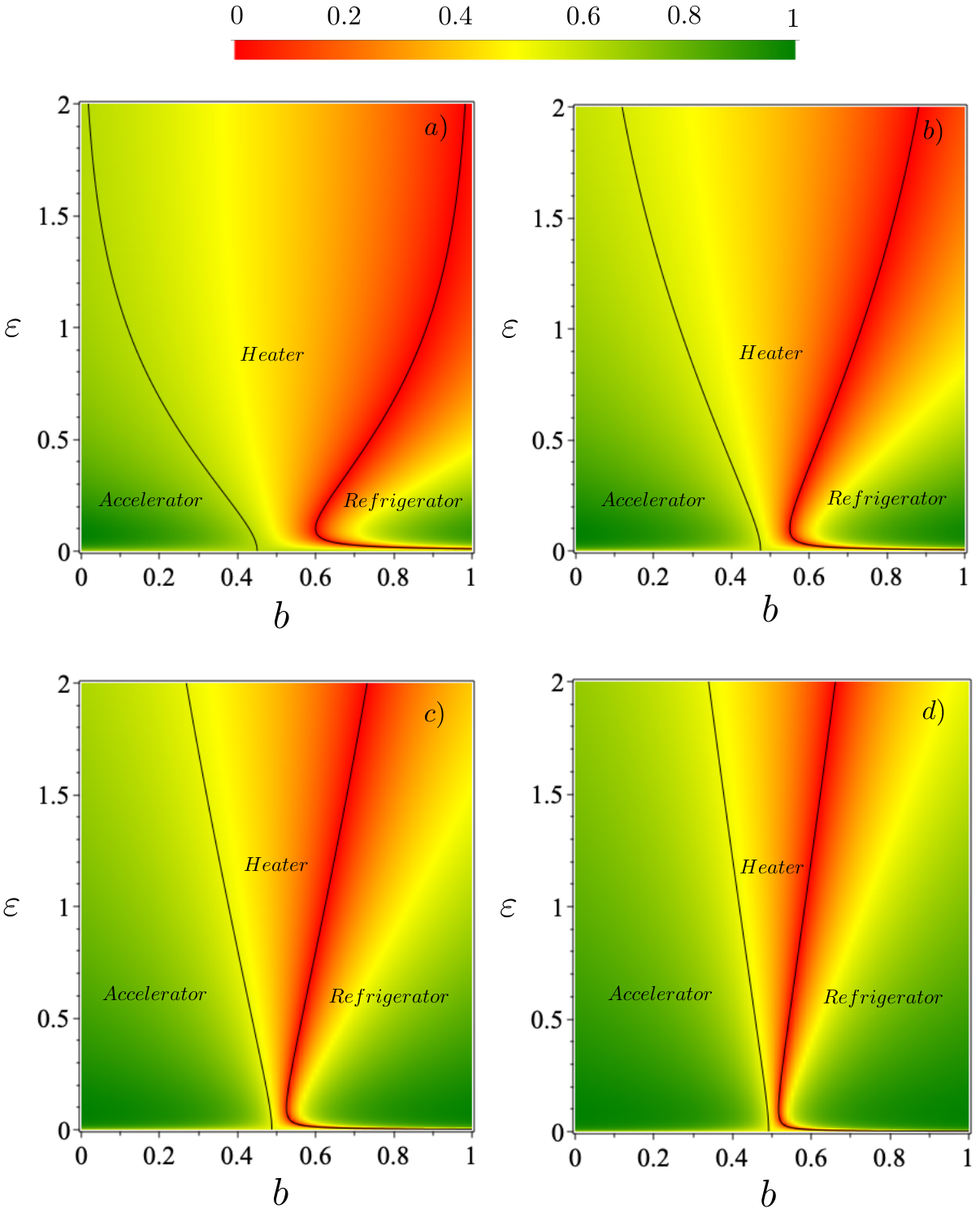} % Substitua pelo nome do arquivo da imagem
    \caption{Thermal efficiency of the quantum cycle as a function of $b$ and $\varepsilon$ for different parameter configurations. a) For  $T=1$. b) For $T=2$. c) For $T=4$. d) For $T=6$. The parameter $\tau$ was fixed at $\tau=0.1$.}
    \label{fig:refri-eff-1}
\end{figure*}

We first consider the limit of vanishing tunneling, $\tau=0$ \cite{vinicius}. Figure \ref{fig:refri-1(zero)} shows the operational modes of the cycle in the $(b,\varepsilon)$ plane for two temperatures. In panel (a), for $T=1$, the accelerator regime occupies the low-$b$ sector, the heater regime dominates the central portion of the diagram, and the refrigerator regime appears for larger values of $b$. In panel (b), for $T=3$, the same three regimes are preserved, but their relative weights change substantially: the refrigerator regime expands over a broader portion of the high-$b$ sector, the heater region is compressed, and the accelerator region becomes more prominent at small $b$. Therefore, in the absence of tunneling, increasing temperature favors both cooling and acceleration while reducing the parameter window associated with heater operation.

Figure \ref{fig:refri-eff-1(zero)} displays the corresponding performance maps in the same parameter space. In the panel (a), for $T=1$, the highest values of the normalized performance $\kappa$ are concentrated near the bottom of the diagram, indicating that the cycle performs best for small detuning $\varepsilon$, both in the accelerator sector $(b<0.5)$ and near the onset of refrigeration for $b>0.5$.  effectiveness of measurement-driven cooling and acceleration. By contrast, the heater region is associated with intermediate or low values of $\kappa$. Overall, Fig. \ref{fig:refri-eff-1(zero)} shows that the best performance is achieved for weak detuning, while increasing temperature enlarges the parameter domain where the cycle operates efficiently.

\begin{figure*}[t] % [t] posiciona a figura no topo da página
    \centering
    \includegraphics[width=1.2\columnwidth]{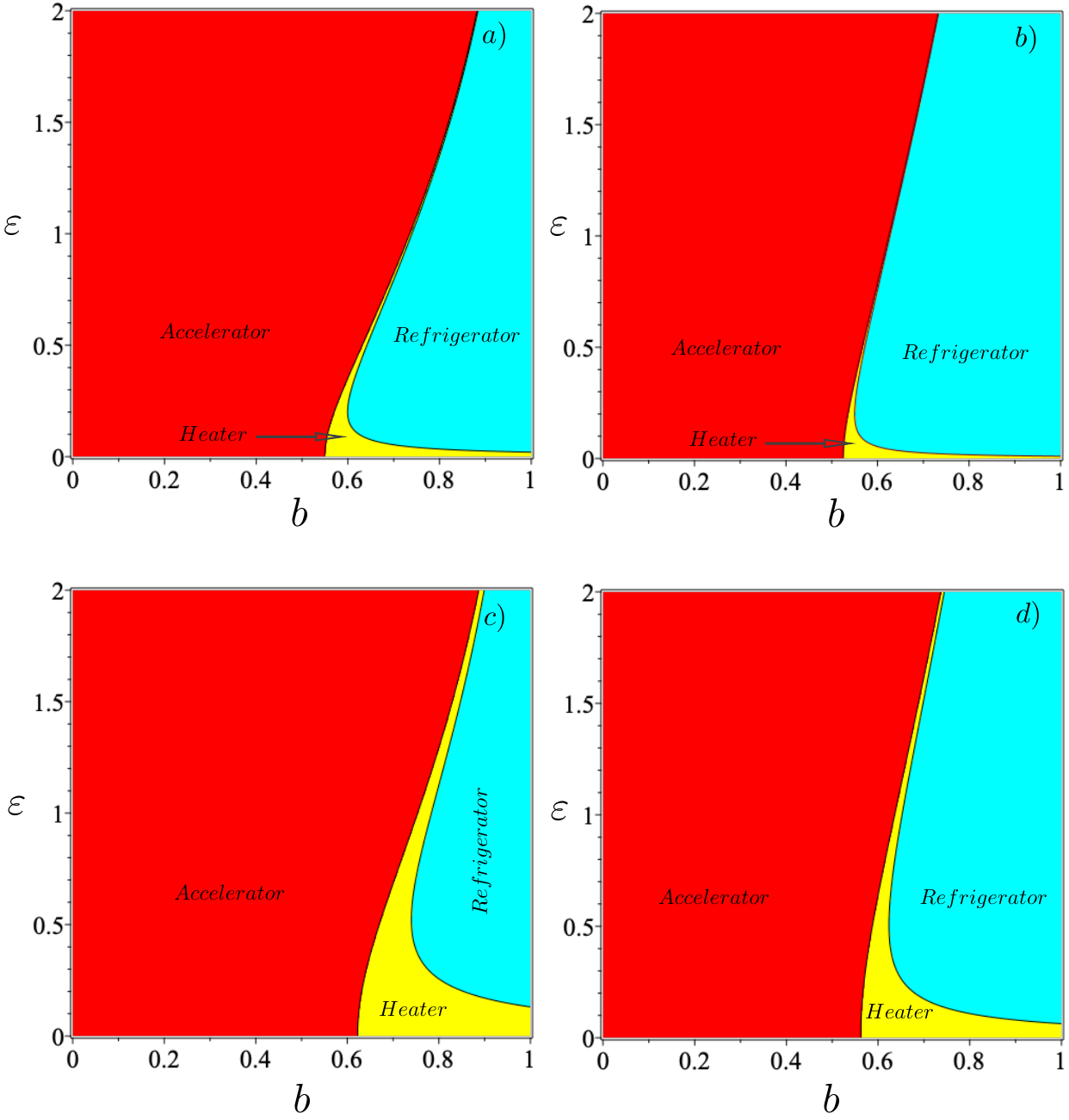} % Substitua pelo nome do arquivo da imagem
    \caption{The operational modes of the quantum cycle as a function of $b$ and $\varepsilon$ for different parameter configurations. a) For  $T=2$, $\tau=0.2$. b) For $T=4$, $\tau=0.2$. c) For $T=2$, $\tau=0.5$. d) For $T=4$, $\tau=0.5$.}
    \label{fig:refri-2}
\end{figure*}

\begin{figure*}[t] % [t] posiciona a figura no topo da página
    \centering
    \includegraphics[width=1.2\columnwidth]{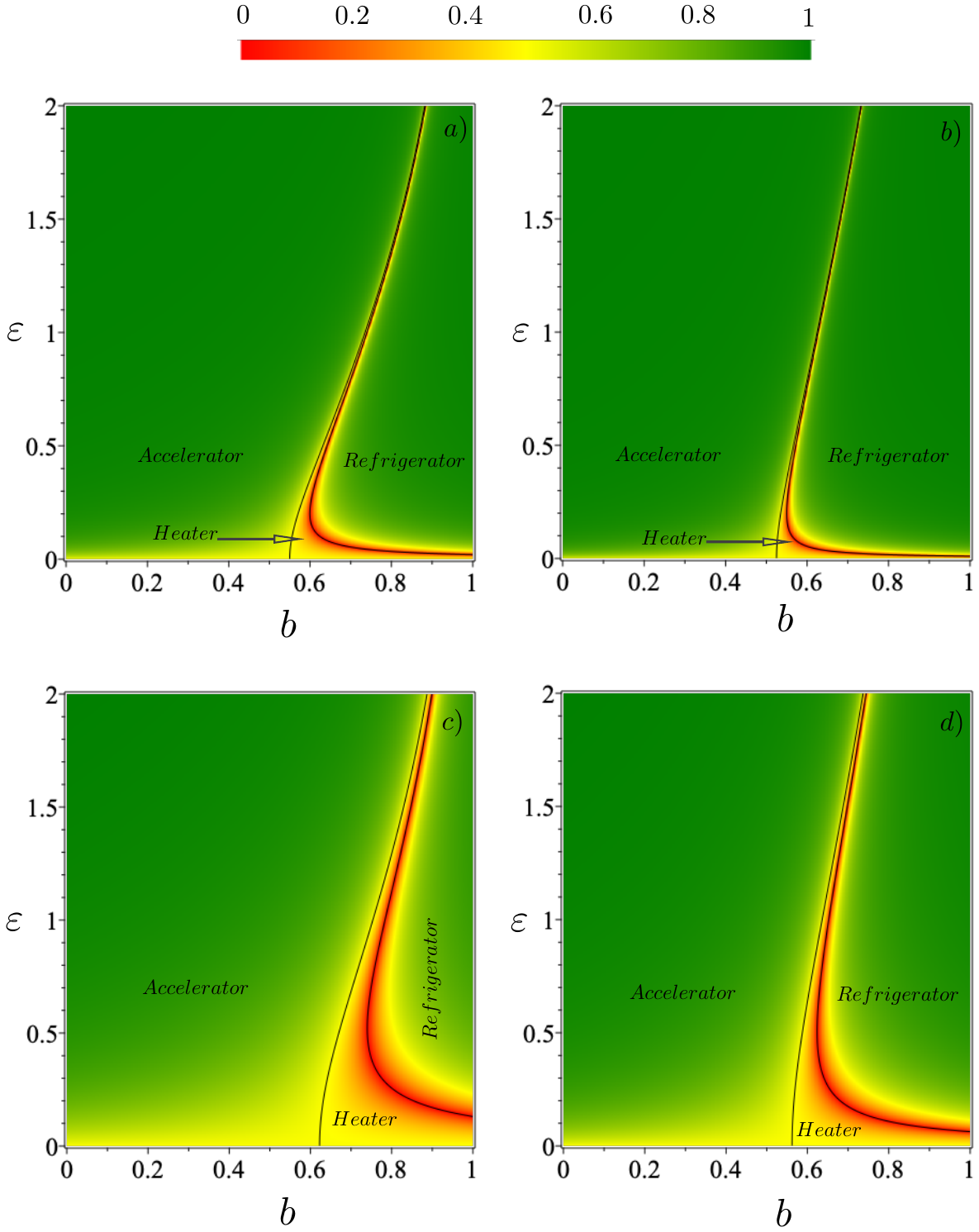} % Substitua pelo nome do arquivo da imagem
    \caption{Thermal efficiency of the quantum cycle as a function of $b$ and $\varepsilon$ for different parameter configurations. a) For  $T=2$, $\tau=0.2$. b) For $T=4$, $\tau=0.2$. c) For $T=2$, $\tau=0.5$. d) For $T=4$, $\tau=0.5$.}
    \label{fig:refri-eff-2}
\end{figure*}

%\begin{figure}
%\includegraphics[scale=0.53]{Fig-ott-TB}

%\caption{\protect\label{fig:Ott-dgm-def}(a) Operation mode of the quantum
%Otto cycle for ${\rm Cu}_{3}$-like compound in the plane $B_{1}-T_{h}$,
%(c-d) Thermal efficiency under the same condition to (a-b) respectively.}
%\end{figure}

We now analyze the effect of the tunneling parameter $\tau$ on the performance of the device. In this configuration, the interplay between tunneling and measurement backaction significantly modifies the operational regimes.

As shown in Fig.  \ref{fig:refri-1} (a), for the lowest temperature considered, $T=1$, the phase diagram is largely dominated by the heating regime, which spans most of the intermediate range of $b$ and extends toward large values of $\varepsilon$. For small values of $b$, the generalized measurements act predominantly as external sources of energy, driving the cycle into the accelerator mode, in which both heat and work are injected into the system and the excess energy is subsequently dissipated during thermalization. Only for sufficiently large values of $b$ does a narrow band corresponding to the refrigeration regime appear, indicating that in this region the back-action of the measurements is oriented so as to promote a net extraction of energy from thermal reservoir throughout the cycle. In Fig. \ref{fig:refri-1} (b), increasing the temperature to $T=2$ leads to a significant change in the operational regime boundaries. The refrigeration becomes broader and shifts toward slightly lower values of $b$ compared to Fig. \ref{fig:refri-1} (a). This behavior stems from the enhanced thermal exchanges with the reservoir, which now assist the measurement back-action in achieving net energy extraction. In contrast, the heater domain becomes narrower, while the accelerator region remains concentrated in the same range of small $b$ values. Thus, raising the bath temperature favors refrigerator operation and reduces the window in which the cycle acts as a heater. In Fig. \ref{fig:refri-1} (c), for the temperature considered $(T=4)$, the refrigeration region undergoes a pronounced expansion, occupying a substancial portion of the diagram for moderate values of $\varepsilon$. In contrast, the heating regime shrinks to a very narrow and highly confined band, indicating that in this thermal range net energy dissipation becomes less likely. The accelerator mode remains restricted to small values of $b$, showing that even at elevated temperatures only very weak measurements are capable of injecting energy into the system in a dominant manner. In Fig. \ref{fig:refri-1} (d), corresponding to the highest temperature analyzed $T=6$, the refrigeration region comes to dominate almost the entire right-hand sector of the diagram, indicating that cooling becomes the predominant operational mode across a wide range of values $b$ and $\varepsilon$. In contrast, the heating regime shrinks to a narrow and well-defined band, while the accelerator mode remains confined to the region of weak measurements, approximately for $b<0.5$.

Figure \ref{fig:refri-eff-1} displays the corresponding thermal-performance maps, considering the tunneling amplitude fixed at the value $\tau=0.1$. In panel (a), for $T=1$, the largest values of the normalized coefficient of performance are concentrated at small detuning $\varepsilon$ and moderate values of the measurement strength $b$, especially close to the onset of the refrigerator regime. This indicates that, at low temperature, efficient cooling requires a careful balance: the second measurement must be strong enough to extract heat from the bath, but weak enough to avoid excessive entropy production. In accelerator sector, located at small values of $b$, the performance is also relatively high when the detuning $\varepsilon$ remains small. This shows that, in the weak-measurement regime, the cycle can still amplify the natural energy flow efficiently, provided that the level detuning is not too large. The heater region, which occupies the central part of the diagram, is instead associated with lower performance, reflecting the fact that a substantial fraction of the injected energy is dissipated as heat rather than converted into a useful thermodynamic task. As $\varepsilon$ increases, the performance decreases in both the refrigerator and accelerator sectors, showing that large level detuning suppresses the effective thermodynamic response of the cycle. In panel (b), for $T=2$, the high-performance region becomes broader and extends over a wider interval of $b$ and $\varepsilon$, reflecting the same tendency already observed in the operational diagram: increasing temperature favors the refrigerator regime and also improves its efficiency. The accelerator mode remains visible at low $b$ and exhibits good performance especially for small values of $\varepsilon$, whereas the heater domain becomes progressively less relevant both in area and in thermodynamic quality. In the panel (c), corresponding to $T=4$, the region of better performance occupies a substantial portion of the diagram, indicating that the combined action of thermal fluctuating and finite tunneling makes refrigeration more robust throughout parameter space. In this regime, the accelerator sector survives only in a restricted low-$b$ portion, but it still presents its best performance at small detuning, while the heater mode is reduced to a narrow low-performance strip. Finally, in panel (d), for $T=6$, the broad green region shows that the refrigerator not only dominates operationally, but also maintains a relatively high coefficient of performance over an extended range of measurement strengths and detunings. The accelerator regime remains confined to weak measurements, with its most favorable response again concentrated at low $\varepsilon$, and the heater mode becomes marginal, confirming that at high temperature the cycle strongly favors refrigeration over the other two operational behaviors. Overall, Fig. \ref{fig:refri-eff-1} shows that, for finite but weak tunneling, increasing temperature enhances both the extent and the quality of the refrigeration regime, while the accelerator retains good performance in the low-$b$, low-$\varepsilon$ sector and heater mode remains restricted to less efficient regions of the parameter space.

\subsubsection{Branch $a = \frac{1}{2}\left[1 - \tanh(\beta E)\right]$} 
For the particular branch
\begin{equation}
a=\frac{1}{2}\left(1-\tanh\left(\beta E\right)\right ),
\end{equation} 
the heat exchanged with the cold reservoir, the work performed during the measurement stroke, and the heat transferred to the measurement channel $M^{b}$ read 
 \begin{align}
 &Q_c=\left\langle\Delta U_1\right\rangle=-E\tanh{\left(\beta E\right)}+\varepsilon(2b-1). \nonumber \\
&W=\left\langle\Delta U_2\right\rangle=(E-\varepsilon)\tanh{\left(\beta E\right)}.\\ 
&Q_h=\left\langle\Delta U_3\right\rangle=\varepsilon(1+\tanh{\left(\beta E\right)}-2b). \nonumber
\end{align}
For this branch, the work remains positive provided $E>\varepsilon$. Notably, this regime is only meaningful for finite tunneling ($\tau \neq 0$), since in the absence of tunneling the work contribution vanishes identically. Consequently, the operational mode of the device is fully determined by the signs of $Q_{c}$ and $Q_{h}$, which are controlled by the measurement parameter $b$. In this branch, the refrigerator regime is obtained when the system extracts heat from the cold reservoir and releases it during the remaining strokes, leading to the condition $b>\frac{1}{2}\left(1+\frac{E}{\varepsilon}\tanh\left(\beta E\right)\right )$. The accelerator regime occurs for sufficiently weak measurements, namely $b<\frac{1}{2}\left(1+\tanh\left(\beta E\right)\right )$, where the measurement backaction predominantly injects energy into the working substance. Finally, the intermediate interval $\frac{1}{2}\left(1+\tanh\left(\beta E\right)\right )<b<\frac{1}{2}\left(1+\frac{E}{\varepsilon}\tanh\left(\beta E\right)\right )$ corresponds to the heater regime, in which the absorbed work is ultimately degraded into heat.

Figure \ref{fig:refri-2} illustrates how the operational modes of the measurement-driven quantum cycle evolve as the tunneling amplitude increases from the weak-coupling regime. In panel (a), for $T=2$ and $\tau=0.2$, the diagram is mainly divided between the accelerator and refrigerator regimes. The finite tunneling amplitude modifies the eigenstates of the double quantum dot and changes the way measurement backaction redistributes energy throughout the cycle. As a consequence, the transition to the refrigeration regime occurs for intermediate and high values of the measurement strength $b$. In addition, Fig. \ref{fig:refri-2}(a) shows a narrow heater region located close to the boundary between the accelerator and refrigerator sectors, rather than a broad band centered at $b\approx 0.5$. For values slightly above this threshold, and more clearly for $b \gtrsim 0.6$, the heater mode survives only at small detuning $\varepsilon$. This indicates that dissipative operation is restricted to the regime of weak level splitting, where the measurement backaction can still convert the supplied energy predominantly into heat. As $\varepsilon$ increases this heater strip rapidly disappears and the system crosses over to the refrigerator regime, which then occupies the high-$b$ portion of the diagram. In panel (b), for the same tunneling amplitude and higher temperature $T=4$, the refrigeration region further expands and occupies a significant portion of the parameter space, indicating that thermal activation and tunneling jointly enhance heat extraction from the cold reservoir. As a result, the heater regime is compressed, becoming less prominent. A more pronounced reorganization of the modes operational modes is observed for larger tunneling amplitude $\tau=0.5$, as illustrated in panels (c) and (d). In panel (c), corresponding to $T=2$ and $\tau=0.5$, the refrigeration regime remais concentrated in the region of comparatively large measurement  strength $b$, occupying the righ-hand side of the diagram over wide range of detuning $\varepsilon$. The main effect of increasing the tunneling amplitude is to shift the refrigeration domain toward stronger measurements, while simultaneously stabilizing it across an extended interval of $\varepsilon$, provided that the measurement strength is sufficiently high. In parallel, the heater regime survives only as a narrow strip, confined to small values of $\varepsilon$ and located near the boundary between the accelerator and refrigerator sectors. Meanwhile, the accelerator mode continues to dominate most of the low-$b$ region. In panel (d), corresponding to $T=4$ and $\tau=0.5$, the refrigerator regime clearly dominates the right-hand side of the diagram, occupying most of the region associated with large measurement strengths $b$. This shows that, at higher temperature and strong tunneling, the cycle is more likely to operate as a refrigerator over a broad interval of detuning $\varepsilon$. By contrast, the heater regime is compressed into a very thin strip restricted to small values of $\varepsilon$ and located close to the boundary between the accelerator and refrigerator sectors. The accelerator mode, in turn, remains essentially
confined to the low-$b$ region on the left-hand side of the
panel. Overall, Fig. \ref{fig:refri-2} shows that increasing either the temperature or the tunneling amplitude systematically favors refrigeration, while suppressing heater-like operation and confining the accelerator regime to the weak-measurement limit.

Figure  \ref{fig:refri-eff-2} displays the thermal-performance maps corresponding to the same parameter sets shown in Fig. \ref{fig:refri-2}. In panel (a), for $T=2$ and $\tau=0.2$, the highest values of the normalized coefficient of performance are concentrated at large detuning $\varepsilon$, both in the accelerator and refrigerator regimes. In the accelerator regime, these values occur for $b<0.5$, whereas in the refrigerator regime they are associated with large $b$, indicating that the performance increases with the measurement strength. The monotonic increase of the coefficient of performance with $\varepsilon$ arises from the fact that the work remains constant along this branch. In contrast, the heater regime, located at small detuning and near the lower part of the diagram, exhibits the lowest performance. This region is dominated by dissipation, where the supplied energy is predominantly released as heat to the environment rather than being converted into a useful thermodynamic task. As a result, the coefficient of performance remains low, particularly near the boundary with the refrigerator regime. A similar qualitative behavior is observed in panel (b), corresponding to $T=4$ and $\tau=0.2$. The increase in temperature leads to a moderate expansion of the high-performance region of the refrigeration regime, indicating that thermal fluctuations enhance the robustness of the cooling process over a broader range of parameters. In contrast, the accelerator and heater regimes become less pronounced, exhibiting a slight reduction in their operational relevance. In the panel (c), for $T=2$ and $\tau=0.5$, the high-performance region of the accelerator extends over a broader range of detuning, indicating that enhanced hybridization between localized states improves the thermodynamic response of the cycle. In contrast, the refrigerator exhibits a reduction in both performance and operational domain compared to panel (a), showing that coherent hybridization, while beneficial for the accelerator, is detrimental to cooling in this regime. Physically, stronger tunneling promotes coherence-assisted population redistribution, enabling the measurement strokes to exchange energy more efficiently with the working medium.  Finally, in panel (d), for $T=4$ and $\tau=0.5$, the refrigerator region expands significantly and exhibits enhanced performance over a broad range of detuning $\varepsilon$. The accelerator regime, predominantly located at low values of $b$, also extends over a wider range of $\varepsilon$, and attains high performance. In contrast, the heater region is reduced to a narrow, nearly imperceptible band, highlighting the effectiveness of the device as a measurement-assisted quantum energy converter. Altogether, Fig. \ref{fig:refri-eff-2} shows that increasing both temperature and the tunneling amplitude not only enlarges the refrigeration regime but also enhances the coefficient of performance over a substantial portion of the parameter space. These results indicate that coherent interdot coupling acts as a valuable resource for optimizing measurement-driven thermal machines based on double quantum dots.

%%%%%%%%%%%%%%%%%%%%%%%%%%%%%%%%%%%%%%%

\section{Conclusions}

In this work, we investigated measurement-driven quantum thermal machines based on sequential nonselective generalized measurements, taking a double quantum dot as a working substance. In contrast to simplified qubit models with purely diagonal Hamiltonians, the present platform incorporates coherent interdot tunneling, which hybridizes the localized states and  hybridizes localized states, and modifies both
the energy spectrum and the thermodynamic response of
the cycle. This allowed us to examine how measurement
backaction and coherent tunneling jointly determine the
operational regimes of the device.

By formulating the cycle in terms of two tunable nonselective measurement channels and a single thermalization stroke, we identified the corresponding energy and entropy variations and used them to classify the distinct operational modes of the machine. Depending on the measurement parameters, the device can operate as a heat engine, accelerator, heater, or refrigerator. In this sense, generalized measurements act as genuine thermodynamic resources: they inject or extract energy from the working substance and can effectively replace conventional hot reservoirs in the implementation of quantum thermal cycles. 

Our results further showed that the interplay between temperature, detuning, and tunneling amplitude strongly reshapes both the operational diagrams and the associated performance maps. In the engine branch, finite tunneling preserves the work-producing regime while quantitatively modifying the boundaries between engine, heater, and accelerator operation. In the refrigeration branch, the increase of temperature generally enlarges the refrigerator sector and improves its performance, whereas the heater mode becomes progressively suppressed. The performance maps also revealed that the best thermodynamic response depends sensitively on the measurement strength and on the detuning, with high-efficiency regions emerging when measurement backaction is strong enough to sustain the desired task without introducing excessive irreversibility. 

An important point emerging from our analysis is that the introduction of the tunneling parameter generates a new configuration of the operational modes associated with the refrigerator branch. In the model with detuning only, the operational structure is more restricted and the boundaries between refrigerator, heater, and accelerator regimes follow the simpler pattern imposed by the diagonal Hamiltonian. Once interdot tunneling is included, however, coherent hybridization changes the energy balance along the cycle and reorganizes the corresponding phase diagrams, giving rise to refrigeration configurations that are absent in the purely detuned model. Therefore, the tunneling term does not merely deform the previous operational regions: it creates genuinely new refrigeration behavior that exists only when coherent interdot coupling is present. 

A central conclusion of this work is that coherent interdot coupling is not merely a microscopic detail of the platform, but an active resource for the optimization of measurement-powered thermal machines. By controlling the tunneling amplitude, one can significantly alter the balance between useful energy conversion and dissipative losses,  thereby enlarging favorable operational regions in parameter space. Our analysis therefore extends the study of measurement-assisted thermodynamics to a more realistic and experimentally relevant setting, and shows that double quantum dots provide a promising platform for the implementation and control of quantum devices in which measurements play a direct energetic role.

\begin{acknowledgments}
M. Rojas and B. Carvalho acknowledge the financial support from the CNPq and FAPEMIG.
M. Rojas also gratefully acknowledges CNPq for support through Grant No. 311565/2025-5. J. F. G.S. acknowledges CNPq Grant No. 420549/2023-4, Fundect Grant No. 83/026.973/2024, and Federal University of Grande Dourados for support. 
\end{acknowledgments}


\begin{thebibliography}{99}


\bibitem{vin} S. Vinjanampathy, J. Anders, Cont. Phys. \textbf{57}, 545 (2016).

\bibitem{kos} R. Kosloff, Entropy \textbf{15}, 2100 (2013).

\bibitem{camp} S. Campbell \textit{et al.}, Quantum Sci. Technol. \textbf{11}, 012501 (2026).


\bibitem{yi} J. Yi, P. Talkner, Y. W. Kim, Phys. Rev. E  \textbf{96}, 022108 (2017).

\bibitem{felce} D. Felce, V. Vedral, Phys. Rev. Lett. \textbf{125}, 070603 (2020).


\bibitem{santos} Z. Li, S. Su, J. Chen, J. Chen, Jonas F. G. Santos, Phys. Rev. A, \textbf{104}, 062210 (2021).




\bibitem{cao} H. Cao, N. -N. Wang, Z. Jia, C. Zhang, Y. Guo, B. -H. Liu, Y. -F. Huang, C. -F. Li, G. -C. Guo, Phys. Rev. Res. \textbf{4}, L032029 (2022).

\bibitem{santos-1} J. F. G. Santos, P. Chattopadhyay, Physica A, \textbf{632}, 129342 (2023).



\bibitem{eb} D. Ebler, S. Salek, G. Chiribella, Phys. Rev. Lett.  \textbf{120}, 120502 (2018).

\bibitem{na} N. Behzadi, Phys. Rev. E, \textbf{111}, 054137 (2025).


\bibitem{def} S. Deffner, S. Campbell, \emph{Quantum Thermodynamics: An Introduction to the Thermodynamics of Quantum Information}. Morgan \& Claypool Publishers (2019). 
 
\bibitem{gem} J. Gemmer, M. Michel, G. Mahler \emph{Quantum Thermodynamics: Emergence of Thermodynamic Behavior Within Composite Quantum Systems}. Springer-Velang  Berlin Heidelberg (2009).


\bibitem{Quan} H. T. Quan, Y. X. Liu, C. P. Sun, F. Nori, Phys. Rev.
E, \textbf{76}, 031105 (2007).

\bibitem{deffner} S. Deffner, Phys. Rev. E, \textbf{113}, 044113 (2026).



\bibitem{myers}N. M. Myers, O. Abah, S. Deffner, AVS Quantum Sci. \textbf{4}, 027101 (2022). 

\bibitem{bata} J. P. S. Peterson, T. B. Batalh\~{a}o, M. Herrera, A. M. Souza, R. S. Sarthour, I. S. Oliveira, R. M. Serra, Phys. Rev. Lett.  \textbf{123}, 240601 (2019).

\bibitem{kasloff}R. Kosloff, J. Chem. Phys. \textbf{80}, 1625 (1984).

\bibitem{alicki}R. Alicki, J. Phys. A: Math. Gen. \textbf{12}, L103
(1979).

\bibitem{mrojas} J. L. Diniz, M. Rojas, C. Filgueiras, Phys. Rev.
E \textbf{104}, 014149 (2021). 

\bibitem{rojas} Onofre Rojas, Moises Rojas, Ann. Phys. (Berlin)
 \textbf{537}, 2400291 (2025). 

\bibitem{mrojas-1} Onofre Rojas, Moises Rojas, S. M. de Souza, Phys. Rev.
E \textbf{111}, 044121 (2025). 


\bibitem{elouard} C. Elouard, D. Herrera-Mart\'{\i}, M. Clusel, A. Auff\`{e}ves, Phys. Rev. Lett. \textbf{118}, 260603 (2017).

\bibitem{klatzow} J. Klatzow, J. N. Becker, P. M. Ledingham, C. Weinzetl, K. T. Kaczmarek, D. J. Saunders, J. Nunn, I. A. Walmsley, R. Uzdin, E. Poem, Phys. Rev. Lett. \textbf{122}, 110601 (2019).

\bibitem{ore} O. Oreshkov, T. A. Brun, Phys. Rev. Lett. \textbf{95}, 110409 (2005).

\bibitem{die} P. R. Dieguez, A. Angelo, Phys. Rev. A \textbf{97}, 022107 (2018).

\bibitem{be} N. Behzadi, J. Phys. A: Math. Theor. \textbf{54}, 015304 (2021).

\bibitem{die-1} V. F. Lisboa, P. R. Dieguez, J. R. Guimar\~{a}es, J. F. G. Santos, R. M. Serra, Phys. Rev. A \textbf{106}, 022436 (2022).


\bibitem{brun} O. Oreshkov, T. A. Brun, Phys. Rev. Lett. \textbf{95}, 110409 (2005).




\bibitem{bran} K. Brandner, M. Brauer, M. T. Schmid, U. Seifert, New J. Phys. \textbf{17}, 065006 (2015).

\bibitem{aro} A. Z. Goldberg, K. Heshami, L. L. Sanchez-Soto, Phys. Rev. Research \textbf{5}, 033198 (2023).

\bibitem{cha} Ch. Purkait, A. Biswas, Phys. Rev. E \textbf{107}, 054110 (2023).

\bibitem{zh} T. Zhang, H. Yang, S-M Fei, Phys. Rev. A \textbf{109}, 042424 (2024).

\bibitem{fra} G. Francica, J. Goold, F. Plastina, M. Paternostro, npj Quantum Inf.  \textbf{3}, 12 (2017).

\bibitem{and} A. H. A. Malavazi, R. Sagar, B. Ahmadi, P. R. Dieguez, PRX Energy \textbf{4}, 023011 (2025).


\bibitem{haya} T. Hayashi, T. Fujisawa, H. D. Cheong, Y. H. Jeong, Y. Hirayama, Phys. Rev. Lett. \textbf{91}, 226804 (2003).

\bibitem{mr} C. Filgueiras, O. Rojas, M. Rojas, Ann. Phys. (Berlin) \textbf{532}, 2000207 (2020).

\bibitem{mr-1} Z. Dahbi, M. F. Anka, M. Mansour, M. Rojas, C. Cruz, Ann. Phys. (Berlin) \textbf{535}, 2200537 (2022).

\bibitem{mr-2} V. Leit\~{a}o, O. Rojas, M. Rojas, Phys. Rev. A \textbf{112}, 022401 (2025).







\bibitem{vinicius} P. R. Dieguez, V. F. Lisboa, R. M. Serra,
Phys. Rev. A \textbf{107}, 012423 (2023).

\bibitem{vinicius-1} V. F. Lisboa, P. R. Dieguez, K. Simonov, R. M. Serra,
Quantum Sci. Technol. \textbf{11}, 015058 (2026).




%\bibitem{X.-L.Huang}X.-L. Huang, X.-Y. Viu, X.-M. Viu, X.-X. Yi,
%Eur. Phys. J. D \textbf{68}, 32 (2014)



%\bibitem{choi08}K.-Y. Choi, N. S. Dalal, A. P. Reyes, P.L. Kuhns,
%Y. H. Matsuda , H. Hojiri, S. S. Mal and U. Kortz, Phys. Rev. B \textbf{77},
%024406 (2008)








\end{thebibliography}
\end{document}